# How to measure musical preference on Facebook

*Evidence from a mixed-method data collection*

Zoltán Kmetty[1], Renáta Németh[2]

1: Eötvös Loránd University of Budapest, Faculty of Social Sciences. Assistant professor, kmetty.zoltan@tatk.elte.hu and Centre for Social Sciences – MTA Centre of Excellence, CSS-Recens Research Group,

2: Eötvös Loránd University of Budapest, Faculty of Social Sciences. Associate professor, Head of Department of Statistics, nemethr@tatk.elte.hu

**Corresponding author**

Zoltán Kmetty, kmetty.zoltan@tatk.elte.hu. Eötvös Loránd University, Faculty of Social Sciences, 1117, Budapest, Pázmány Péter sétány 1/A, Hungary

**Funding**

This research was supported by the National Research, Development and Innovation Office of Hungary (NKFIH), grant number FK-128981.

**Abstract**
More and more digital data is available for social science analysis. This provides new ways of measuring several concepts. But when we start using new data sources, we have to understand how the new data source could be processed and how it could be analysed effectively. It is especially for Facebook data since there is no established gold standard analysis-framework. However, researchers have in-depth knowledge on how to measure different concepts using survey data. Thus, cross-referencing Facebook data with survey data is a reasonable way to support Facebook data analysis at different decision points. In this paper, we present how music preference could be measured by Facebook data and how survey data could support the selection of main indicators. Based on our results, we provide some general suggestions for Facebook data processing and indicator operationalization.

**Introduction**

A valid measurement of concepts is a key issue of social science. The validity of surveys has been criticized for a long time because they create an artificial environment while collecting data with some pre-specified purpose. In contrast, digital data like Facebook (FB) yields 'organic data,' that is, observational data of users' behaviors. However, validity is a concern in the case of FB, too. The biases introduced by shifts in algorithms, platform usage changes, or fake self-representation conformed to social expectations raise possible validity issues. However, the biggest challenge that digital data present is indicator operationalization. Researchers have more "space" to create their own indicators, and there is no gold standard on how to measure certain things. With so many operationalization options, it is very hard to speak about "valid" measurement.

In contrast, quantitative social scientists have in-depth knowledge on how to measure different concepts using survey data. We know how to create good questionnaires, we know how to handle missing data, and we know how to measure certain concepts. We can develop indexes or factors based on our well-chosen and pre-selected survey items. But when we have social media data, we have to create our variables from scratch.

Our study investigates whether there are transition paths between survey and social media (Facebook) data and whether we can assist the Facebook data-based operationalization of certain concepts by survey data. In this study, we use a parallel mixed-method data collection: a face-to-face survey combined with personal FB data archive collection. The topic we chose here to measure is music preference, a key indicator in cultural sociology, and whose 'digital trace' has its own relevance since the Internet has been the primary source of music consumption.

There were several API based Facebook data collections before the API desert (Freelon 2018), but the use of the personal FB data archive is very novel in scholarly research. We found only four previously published examples (Astudillo et al., 2018, Eslami et al., 2018, Marino et al., 2017, and Thorson et al., 2019), among which only the last one pertains to social sciences. The personal data archive provides access to different Facebook activities, but the effective process of these data sources is not discussed in the academic literature. Our paper is a first attempt to provide general solutions on indicator operationalization within this data environment.

**Internal validity of data - known problems in different types of datasets**

Internal validity (subsequently 'validity') of data depends on the accuracy of the measurement, on whether we actually measured what we were supposed to measure. It is a basic concept of survey methodology (see, e.g., Lavrakas, 2008). Some of the most relevant validity problems of surveys are recall bias, context effect (wording of the question, etc.), interviewer-related measurement error, etc. Validity (moreover, objectivity) of surveys has been criticised for a long time also from a more general epistemological point of view: it has been argued that survey questions are simplistic, incapable of grabbing the complexity of social reality, and are actually a construction of the researchers' own views (e.g., Potter and Shaw, 2018).

FB data, being not self-reported but observational, do not have these issues. However, they are biased for other reasons: shifts in the algorithm (Kosinski et al., 2015, Vitak 2017) or changes in user behavior (e.g., young users tend to leave FB, Perrin, 2018). Moreover, general epistemological issues arise here as well, even because of its observational nature: we have to interpret behavior without the opportunity to ask questions from the observed users. We need some measures that make the unstructured flow of data analyzable, but the measuring instruments are usually not validated with respect to some gold standard. A good example is textual sentiment scores aggregated over documents that aim at measuring happiness.

Finally, there are issues present in both data sources. Social desirability bias, also known in survey research, affects users' self-representation on FB (Kosinski et al., 2015, Gil-Or et al., 2018).

Surveys combined with FB data at an individual level are not only richer but also more robust. Self-reported and observational data on the same phenomena give us the opportunity for validation. The most common method for validating FB data is through independently conducted surveys when aggregated data are compared to each other (see, e.g., Zhan et al., 2019, on a health behavior study). Individually-linked survey and FB data are more accurate (although more expensive and technically more difficult) solutions for validation. In this study, we cross-validate the musical preference of respondents using both survey and Facebook data.

**Musical preference**

Musical preference plays a central role in cultural sociology. Bourdieu (1983) used it as one of the key indicators of cultural consumption in his homology thesis. Peterson used musical preferences to study the omnivorization process in culture (Peterson 1992, Peterson and Kern 1996), and DiMaggio (1982) used music to measure cultural capital in his classic work. The big advantage of musical preference compared to other cultural indicators is that it does not depend heavily on the household's financial situation; everyone can have access every type of music.

In survey studies, musical preferences are typically determined by asking respondents to rate a list of 10-15 genres, using Likert scales to express how much they like them (the most widely used questionnaires were developed by Rentfrow and Gosling, 2003, and Sikkema, 1999). There are other measurement approaches like the artist-based or content-based (audio) approach or, in non-survey settings, the use of music streaming data. A common point of these methods is to indirectly obtain the respondent's preferred music genres because it is not obvious that people have the same understanding of what the genre categories mean (Park et al. 2015) and what are the boundaries between these categories (Beer and Taylor 2013). Indeed, in a survey, it is also possible to ask respondents to evaluate certain bands (Ferrer et al. 2013) or listen to specific songs, but these approaches also induce several validity questions. Why are these bands/songs are chosen, are they a good representatives of a genre, are they recognizable by participants, how many bands/songs are needed to measure a dimension, etc.

As the Internet plays a crucial role nowadays in music listening, the measurement of musical preference has become possible through digital traces (Greenberg and Rentfrow 2017). Several online platforms have been used to measure musical preference, like Twitter (Park et al. 2015), LastFm (Berkers 2012), or Facebook (Nave et al. 2017). Nevertheless, as far as we know, there are no studies on the relationship between digital trace data and survey data regarding musical preference. Our study is the first step in this field.

**Data**

Our study uses a novel joint data source of combined Facebook and survey data. As Facebook restricted the use of their API to access a large number of Facebook data (Freelon 2019), new methods have to be developed by social scientists on how to access FB data for social science research.

There are plenty of studies, which concentrate only on users with public pages. Public pages are appropriate for some specific research questions, for example, to study politicians and parties (e.g., Caton et al., 2015). However, even if one extends the scope on the comments sections of these pages, public sites show only a narrow slice of FB. Therefore the new generation of studies collects the data through the users, not the tech company (Halavais 2019). A possible solution is to use applications or browser plug-ins that participants must install (e.g., Haim, Nienierza, 2019). Using this method, researchers can follow what the participants are exposed to (e.g., content of their timeline) but do not see the participant's reaction. Finally, one can solve the accessibility problem by asking participants to download a copy of their FB profile archive (which option is provided by FB) and to give the researchers that part of these summaries, which are in line with legal regulations. Contrary to the previous case, this way, we do not have data on the content the user reacted to, but we have access to her/his reactions/comments. Furthermore, we have data regarding also the past behavior of the participants. In our study, we followed this way of data collection.

The last two solutions require the researcher to contact the study participants in some way, which induces further technical challenges and needs resources. However, this procedure has benefits as well: researchers can introduce a proper sampling strategy when choosing whom to contact (which may result in higher reliability). Additionally, contacting the participants gives the opportunity to ask them for further information. This approach allows the combination of FB data with a classic survey. In our study, merged data contains both self-reported information and online digital traces, which creates a rich dataset of users' attitudes and behavior. Our study is one of the first attempts using this new track in computational social sciences.

One hundred fifty respondents took part in our study. Personal FB data archives are rich enough to conduct a study with even only a single person. Considering our research questions, our sample size is adequate to identify patterns in the relationship between Facebook and survey data.

The basic demographic characteristics of the sample are available in the appendix Table A1. The sample is non-probability quota sample, with quotas for age category and gender. All the respondents were

Hungarian, living in the eastern part of the country, mostly in big cities. The fieldwork was done in 2019 between April and September, by a professional market research company. Participants who agreed to take part in the study were sent a project description and were invited to come to the office of the market research company. Before starting the study, participants were asked to sign an informed consent form. Importantly, preserving participants' privacy was a key issue in this data collection process. To ensure this, the raw data were anonymized right after the export. An R script that replaced all person names with hashed IDs was run by the interviewer while the respondent was present (this is similar to the approach employed by Mancosu and Vegetti, 2020). The raw Facebook data was deleted after the anonymization process. The market research company only shared the anonymized data with the principal investigator of the study.

. We did not ask for access to the full Facebook activity data because of privacy issues. Our database does not contain private messages, search histories, and marketplace activities. Still, the data covers a wide range of Facebook activities: posts, comments, likes and reactions, pages, friends, profile, and ads data. The full friend list contains 116 000 names (anonymized), there are 83 000 page-like from more than 50 000 unique pages, and the database contains more than 1 800 000 reactions as well as all posts and comments of the participants. The data covers the whole time period of the participants' Facebook usage.In this paper, we are using the page likes and the ads data from the archives.

Besides sharing their Facebook data, participants had to fill out an online questionnaire. Questions about politics, media usage, self-representation, mental health, spare-time activities, and musical preferences were asked from the participants.

As we have a convenience sample, we cannot investigate reliability or external validity of our statistics. However, their internal validity can be assessed through a comparison of survey and FB data.

**Indicators and basic statistics**

In our survey study, we decided to use the genre-based approach (for other studies using the same approach see eg. Colley, 2008, George at al, 2007, Delsing et al, 2008 and Schäfer, 2009). We measured nine music genres (classical music, electronic music, hip-hop, jazz/blues, pop, rap, rock, world-music and 'mulatós'), using a 1-7 Likert scale (one means absolutely do not like the genre, and seven means really like it). We decided to develop our own list of genres, and not to use a standard list like the STOMP-R with 23 genres (Rentfrow and Gosling, 2003), because we aimed at adapting the list to the Hungarian local context (see 'mulatós', a special Hungarian genre) and also, we kept the number of genres rather low to make them broad enough to be easily reproducible from Facebook data. Finally, the genres were chosen to be easily recognizable to respondents and to be of current validity.

In the sample, Pop was the most popular music genre, and Classical music got the lowest score (see Table 1). There was one special Hungarian music category, called 'mulatós' (no recognized translation exists). It is a mixture of Roma wedding music and techno. Listening to 'mulatós' correlates with lower socio-economic status (Kristóf and Kmetty 2019). Many of the bands and singers of this music genre belong to the Roma minority in Hungary. The Hungarian elite does not regard 'mulatós' as something of value, stigmatizing its consumers as having 'bad taste,' failing to understand its role and context (Kovalcsik 2010).

**INSERT TABLE 1 HERE**

Table 1. Self-reported music genre preferences in the survey

We decided to create a joint category for the hip-hop and rap genres. Our Facebook page categorization is based on bands (see later in detail), and it is not easy to decide in some cases if a band is a hip-hop band or a rap one. The correlation between the two survey items is 0.7, which also indicates a strong link between the corresponding categories. We calculated the average of the variables. The mean value was 4,1 with a 0,14 standard error.

The Facebook page like data contains two types of information, the date of the page like, and the name of the Facebook page. However, in the page's profile data, a categorization of the given page is also available, which is created by Facebook. The 150 respondents liked 52 701 pages. Three thousand eight hundred three pages were categorized as music, which is higher than 7 percent. If we add that 80 percent of the page categorization is other or unknown, this value turns out to be quite high. We focus on these pages in our study. Two BA students - who study sociology - categorized these 3803 pages manually. Every page was coded by one coder. We use the same genre typology what we had in the survey, so we have distinguished 8 genres (hip-hop and rap in a joint category), and we had another category for genres like Latin or soul. There were 455 additional pages, which we were unable to put in any genres. Festivals and radio stations were in this group and also some pages, which did not relate to music in any way (false FB-classification).

**INSERT TABLE 2 HERE**

Table 2. Classification of the liked Facebook pages classified as music (N=3803)

The three most frequent music genres were rock, pop, and electronic music. There were relatively few pages being categorized as classical music, jazz/blues, world- music, or 'mulatós'.

We summed up the number of page likes per music genre for each respondent and merged it with the survey data. Within the 150 respondents, 11.3% did not have any music page likes. The mean of the music page likes was 39, the median was 13, and the standard deviation was 110. The distribution was highly skewed; one of the respondent's had 1137 page likes. We will analyse this variable in more detail in the next section. We also calculated a dummy variable for all the genres. We coded 0, and 1 page likes into 'not linked to the genre' and all values above 1 into 'linked to the genre'. In the following, we call this variable 'digitally expressed interest'. It is not obvious which cut-off point to choose to define digitally expressed interest. We decided to code 1 page like into 'not linked to the genre' because the genre boundaries are not clear cut, and we can assume that our coders might have put bands to different categories from where respondents would have classified them.

**INSERT TABLE 3 HERE**

Table 3. Music genre page likes on respondent level

The number of musicians is different in every music genre, and this has an effect on the mean value. There are more pop bands than classical music artists. So the mean value alone is misleading. The dummy variable presents a more valid picture of the musical preferences. Based on that - pop music is the most preferred genre, and hip-hop/rap category is closer to the leading genres, compared to the analysis based on the mean value.

The third data source we use in this paper is the ads interest data. Facebook categorizes every user for sales for advertising. This is an algorithmic classification of the users based on their own likes, activities, and used keywords and also based on their friends' preferences (DeVito 2017). The algorithm is a black-box; we can only observe the result of the categorization. Thorson et al. (2019) have proved that political news exposure is strongly correlated with the ads categories users are classified.

Overall there were 18 689 ad interest categories in the sample. There are high-level categories like jazz music, more specified subcategories like British hip-hop, and specific musicians and bands, too, like Eminem or Adele. We used the high-level categories as they contain all the lower-level categories. So if someone is classified in the British hip-hop category, he/she would also be categorized in the hip-hop category. The list of used interest category names is available in the appendix (table A2). We could extract 7 music genres. We didn't find any interest group for the 'mulatós' preference. This reveals a very important problem. The category selection of Facebook limits the measurability of certain interest groups. This is especially true for categories not typical in the main Facebook countries (primarily the US).

We aggregated the ads interest data - called "inferred interest" from here on - to respondent level and merged it with the survey. Seven percent of the respondents were not linked to any music category through inferred interests, and 5 percent were assigned to all music genres. An average respondent was linked to 3.8 genres. Eighty-one percent of the respondents were assigned to rock music, 68-68 percent to pop and electronic music, 73 percent to hip-hop/rap, 32 to jazz, 30 to world- music, and 24 percent to classical music.

There are two other data sources of Facebook archives that could be used for analysing musical preferences. One of them is the reactions of the users. If they like a post on a page of a musician, we can map this activity. However, the explanatory analysis of this dataset showed that only around half of the respondents used any reactions at least ten times. So this dataset tells us more about people's Facebook usage habits, than their musical preferences.

The user's post database also contains information about musical preferences. In their posts, people can share what music they like, or what music they listen to. A typical form for this, if they share what music they listen through a music streaming applications like Spotify or Deezer (it is a Hungarian one). Nevertheless, the preliminary analysis of posts data revealed that this kind of Facebook usage is rare, and we cannot map the musical preferences of our sample through the user's post data.

**Digitally expressed interest**

Users without music page like

People are using Facebook in several ways. If we operationalize musical preferences through page likes, we have to assume that users follow their favorite musicians and bands. Eleven percent of the sample did not have any music-related page likes. It is an important validity question why don't they have any page likes. One possible explanation is that they do not have any interest in music, and this behavior is expressed in their (non)-activity. However, it is also possible that they don't use this functionality of Facebook, and/or they are just using Facebook lightly. We tested all of these explanations. First, we calculated the number of really liked genres based on self-reported interest data (6 or 7 on the scale) and analyzed the average of this variable within those who had music page like and within those who didn't. There was a small, and not significant ($p=0.33$) difference between the two groups (1.5 vs. 1.9). Second, we run a crosstab analysis to measure the relationship with Facebook activity. Seventy-five percent of the whole sample uses Facebook more than once a day. This proportion is 70 percent within those who don't follow any music-related page, and 76 percent within those who follow at least one page, so the difference here is also small and not significant. Third, we calculated the mean number of all page likes in the two groups. The mean value was 55 within the group with no music page like, and 620 in the other group and this difference was significant (p=0.00) even in this small sample. To sum up the results, based on the three calculations, it seems that people with no music page like have strong musical preference, they use Facebook actively, but they don't use FB page like function.

Correlation of self-reported interest and digitally expressed interest

If we assume that we can measure music preference through Facebook data, we expect to find a modest correlation between digitally expressed interest and self-reported survey interest. In the previous section, we introduced the basic distribution of the variables. The next figure presents the interest order position of the music genres from self-reported interest data and digitally expressed interest data. When defining the latter, we use the proportion of those who like at least two pages per genre.

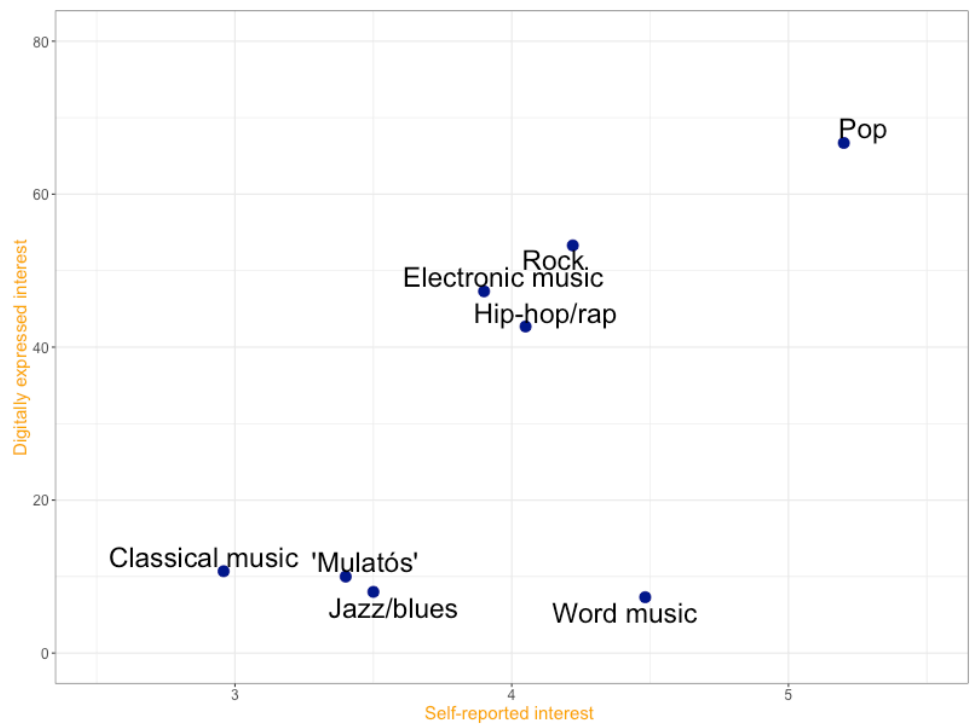

Figure 1. Comparing self-reported interest and digitally expressed interest

Most of the genres are in a similar position in the two approaches. 'Mulatós' and Jazz are evaluated higher in self-reported data, while rock, electronic music, and hip-hop/rap categories are a little bit lower. There is one significant difference, the "world music" category. People like this genre based on the survey, but we can't measure this in the digitally expressed interest data. We calculated the (Spearman's rank) correlation of the survey items and the number of page likes per music genre.

INSERT TABLE 4 HERE

Table 4. Correlation of self-reported interest data (row) and digitally expressed interest data (column)

The diagonal contains the correlation with the same genre. In most cases, the correlation is highest between the same genres, but there are some important deviations. The survey world-music preference does not correlate with the same FB item. We could have expected that based on the difference of the base statistics. The survey answer correlates with the electronic music page likes and the hip-hop/rap page likes. The world-music page like number correlates with the rock survey item. Also, an interesting result, that 'mulatós' survey item correlates with hip-hop/rap stronger than its own category. Overall the correlations are rather moderate; the highest value is 0.44

**Minimum time period to obtain musical preferences**

When we decide to use Facebook likes to measure a concept, we have to decide the time frame of the used data. Facebook archives data provides a timestamp for every user activity. Two questions arise here. Which is the minimum time period of Facebook usage which provides enough data to measure musical preferences. And Is it better to use a smaller, more recent activity data or better to use the whole available time period? ., It is not evident how people use Facebook - do they start to follow/like all the music pages they prefer when they start using the platform or is this a slower process?

To measure the temporal dynamic of musical preference, we created two complex variables. As we wanted to analyse a longer time period of musical preference development, we omit those respondents, who started to use Facebook after 2012 - altogether 26 participants. In the second step, we selected the first seven years of Facebook activity of all users. As most of the users in our sample started to use Facebook between 2010 and 2012, we can assume that possible algorithmic changes do not cause too much bias in this data. Then we calculated for all the respondents the cumulative percent of their Facebook page likes within the elapsed time. Furthermore, we also calculated the cumulative percent of musical preference growth. So if someone preferred two genres within this whole seven-year period, but only one genre after the first year, the first-year cumulative percent was 0.5 in their case. We only include those in this analysis who had at least five music pages likes (N=85).

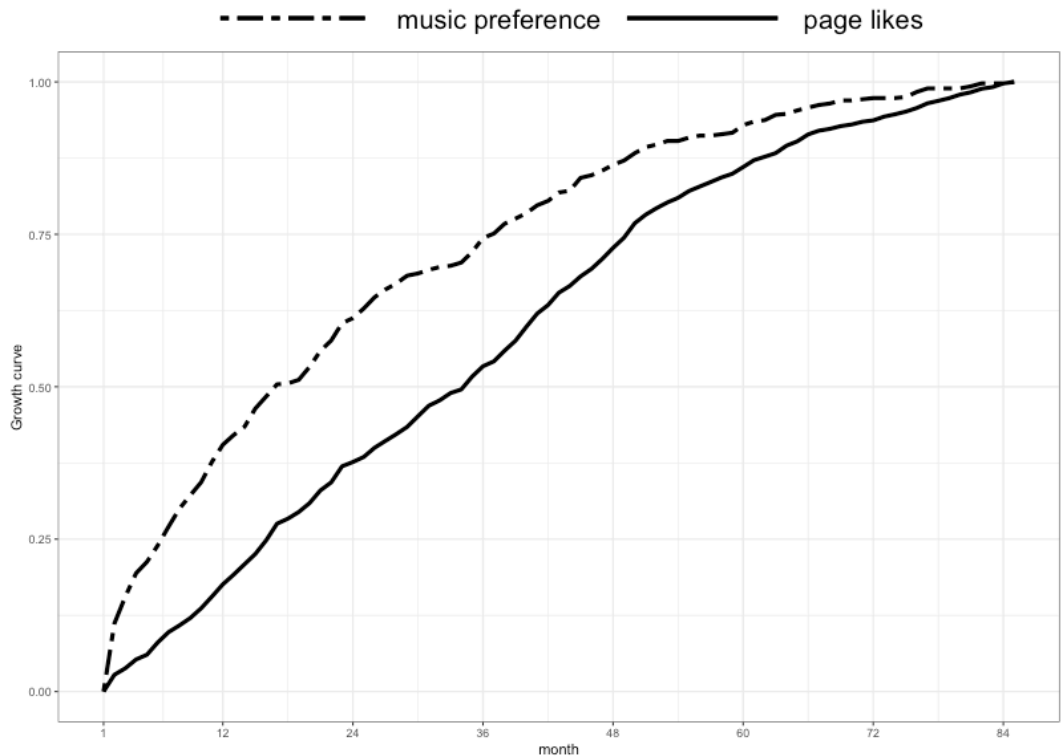

Figure 2. The Cumulative growth of Facebook page likes and ratio of preferred music genres

The blue line represents the growth of page likes. This figure has at least two important messages. At the start of FB use, people will not start to follow all their preferred musicians. So we cannot observe a strong kick-off increase here. Additionally, the growth of page likes do not show a linear trend for the whole period, but for the first 5 years, the trend is linear. The increase has a saturation point around year five. So after a linear growth, the growth rate is starting to decrease. This trend could be a general one but could also be caused by the age distribution of the sample as it is biased toward young users, so this could be an age effect too.

The growth rate of musical preference growth presents a different pattern, and it is far from linear. After two years of Facebook usage, we can map 61 percent of the genres a user likes, after 3 years, 74 percent, and after 4 years, 86 percent. We could assume that the growth rate is even steeper for older generations where musical preference is more stable.

Based on the two trends, we can expect that the current musical preference measured by the survey would correlate stronger with the "whole time period" Facebook page like data, than with the last few years of activity. To measure this, we calculated the number of page-likes per genre from 2018 and then correlated these variables (Spearman correlation) with the survey responses (1-7 scale). This is the same analysis as we did for the whole time period in the previous section (see table 4). The correlation was lower in all cases between the same genres. In some cases (pop), the correlation turned from positive to negative.

These results have important validity aspects. If we want to analyse someone musical preferences through Facebook page likes (digitally expressed interest data), the preferred solution is to use the whole timeline of the user, because a few year time-frame could be misleading, especially if we just use the last few years of activity, and don't include the first (5) years of Facebook presence. We also have to omit those users who have too short Facebook presence, as their musical preference measurement is biased based on the survey validation.

**Inferred interest**

The last part of the analysis focuses on the ads interest data and its correlation with the other measurement methods. The ads classification algorithm is a black box, but some available high-level patents give us some idea about how they work. The algorithm uses the page likes, search words, comments and shares, and even the interest of friends.

To our knowledge, no previous study has reported the rate at which any population is categorized by Facebook to advertisers as interested in music genres. Inferred interest overall shows a stronger preference level than the other measures. Eighty-one percent of our respondents were linked to rock music, and even the least popular classical music was linked to every fourth of our respondents. The next figure (3) gives an overview of the difference between the survey data and the inferred interest data.

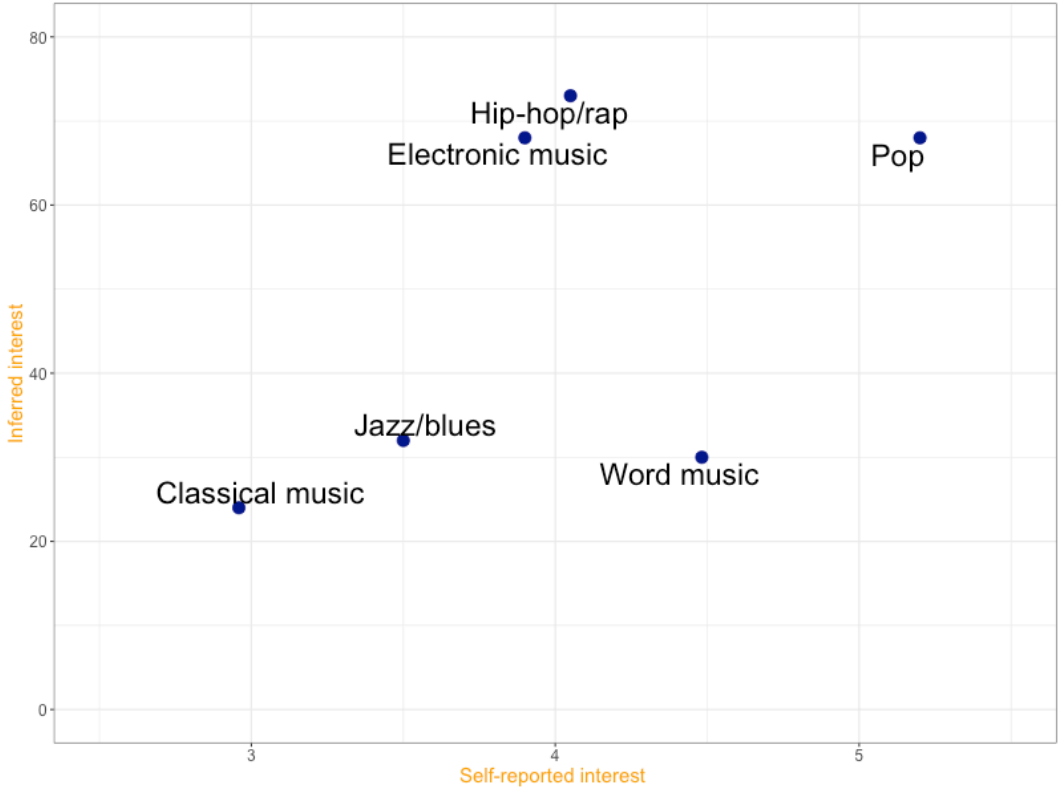

Figure 3. Comparing self-reported interest and inferred interest

The basic tendency is the same, as in the case of page likes. Rock, electronic music, and hip-hop/rap genres were over-measured in the inferred interest data, compared with the self-reported interest. World-music position is completely different here, too; it is preferred much more in the survey. We also have to highlight that the difference between the preferences of genres was bigger here than in the case of the page likes. We expected the opposite before the analysis. It is possible that the algorithm overestimates the homophily of musical preference, and this might cause a bias in this data.

**Discussion**

Our study contributes to the research on social media data by combining personal FB data archives with self-reported data. Our methodological results add to the technical feasibility of such studies, while our substantive results provide important conclusions on how to create valid indicators using Facebook data.

This paper is the first attempt to cross-validate the self-reported musical preference with Facebook-based musical preference classification. We primarily focused on operationalization-related questions and validity issues. The results overall showed that the different measures have only moderate correlations with each other. Some genres measured similarly, but there were significant differences too. A good example of the latter is world-music. It was the second most preferred genre of the survey, but based on FB data (both on digitally expressed and inferred data), it was at the lower end of the preference scale. We can assume that this music category is too broad, and people might disagree on what belongs here. However, it is also possible that they love this genre, but they can't express their preference because the genre is under-represented on Facebook. We know that the affordance of the platform and the algorithms forming the timeline can also affect the (expressed) interest of users. However, the classification of musicians/bands into genres is far from obvious in some cases. A musician could have songs in different genres, and the genre boundaries are not clear in some cases. That is why we merged the hip-hop and rap categories.

It seems that the size of a genre (how many musicians/bands are in it) also affects the operationalization. Furthermore, we have to take into consideration that different genres could have different representation on FB. If a musician uses different platforms other than Facebook (like Soundcloud or Instagram), it might cause a blind spot for our measurement. Thus, the number of likes per genre might not be the best indicator to measure musical preference. For this reason, we created a dummy variable, but this has two disadvantages. First, we lost the scale character of the original variable, which is a significant information loss. Second, we have to define a cut-off value. We chose to categorize 1 page-like into 'not linked to the genre' - to decrease the effect of possible measurement error of page classification. However, there is no gold standard on how to deal with this issue.

Nevertheless, even if we could agree on the cut-off value, we still have to deal with those users who don't have any music page likes. We tested three different hypotheses about the lack of music preference. In the analysis, we concluded that people without music page like have an overall strong musical preference, and they use Facebook actively, but they don't use page-like functionality. This is a possible bias that we have to consider when creating indicators based on Facebook page-likes data.

How we can use the algorithmically inferred data is also a validity question. It is based on Facebook's own classification, which is a total black-box for the research community. We can try to define a higher-level classification by merging lower-level ad interest categories into higher-level ones; however, we would still be exposed to the unknown classification algorithm. The digitally expressed and the inferred data showed a strong correlation with each other (above 0.9) - but this is not a surprise; we know that FB includes the page-likes in its algorithm. However, it was not expected that the self-reported musical preference would correlate stronger with the page-likes than the ads interest. We only have weak assumptions for explaining this. One of the possible explanations is based on the fact that Facebook uses the friends' interests to classify the users (e.g., Thorson 2019). The theoretical background for this decision is the well-known social science phenomenon that friends share the same interest (see the thesis of network homophily (McPherson et al. 2001).

Nevertheless, the homophily level is not the same for all the interest categories, like music, sport, or politics. If the algorithm uses an equal level of homophily factor for all interest groups, it might cause a bias in the classification. Fine-tuning of the algorithm could also result in an over-estimation of musical preference. It is always an open question of which type of error we want to minimize in a classification model. By choosing a more gentle cut-off point, we classify some users into a category that he/she does not like. On the contrary, a strict cut-off point results in failing to classify some user into a music genre which he/she likes. If we want to minimize the latter error (which is completely logical from a business perspective), we tend to overestimate the level of musical preferences. The higher degree of musical preference level in ads data may show indirect evidence for such an assumption.

Another important validity problem arose in the case of the algorithmically inferred data. We did not find any ad interest category which fits the 'mulatós' genre. The category selection of Facebook limits the measurability of certain interest groups. Although big data theoretically makes it possible to reach smaller subpopulations, it is not obvious how we can find measures to analyse these groups if this group is not classified by the Facebook algorithm.

When we choose to use Facebook likes to measure a concept, we have to decide on the time frame of the used data. Our results showed that music page-like increased linearly in the first five years of users' FB-usage, but the growth rate started to decrease afterward. Using only the first three years of page-like data, we were able to estimate quite well the whole range of full-period musical preference of the user. We can assume that the initial time interval long enough to estimate musical preference would be even narrower for older generations where the musical preference is more stable (Way et al. 2019). This result has an important operationalization outcome too. If we want to analyse someone's musical preferences through Facebook page-likes (digitally expressed interest data), the preferred solution is to use the whole timeline or the first 3-5 years of the user's Facebook activity. If we only use the last few years of activity, it could be misleading. We also have to omit those users who didn't have long enough activity data. In the case of music preference, a minimum of 3-5 years of activity is needed.

We can provide some general suggestions for the Facebook data process and indicator operationalization based on our results. We list here the three most important ones based on our analysis.

1. People use Facebook in different ways. When we rely on one Facebook dataset (like page-like), we have to keep in mind that some people might not use that functionality at all. The first step is to check the usage pattern of the selected function and identify those whose data need to be omitted from the analysis.
2. If we use any kind of activity data, this data might be highly skewed. Some people use Facebook frequently; others do not. Thus, we have to normalise our indicators or create more robust categorical variables for more valid measurement.
3. Facebook provides dynamic data in most cases. It is not evident which time frame we need to use for analysis. Our study found that the whole time period provides the best result, and using only the last 1-2 years of activity might be misleading. This is not a universal solution. In other types of analysis different time-frames could work better. Our message here for other researchers is to always think about this issue and make a conscious decision on the time-frame, rather than passively use the data which is available.

Digital data - like social media data or search engine data - opens the possibility for examining topics we could not examine before or re-examine topics with new approaches. However, all the data sources have their own validity problems. Platform affordance and shifts in algorithmical classification are known factors causing issues in Facebook-based measurements. In this paper, we discussed another factor: different types of platform usage, different activity patterns. We presented important consequences based on this factor, which influence users' expressed (observable) interest. We need to consider all these factors when relying on social media data.

**Appendix**

INSERT TABLE A1 HERE

Table A1. Basic socio-demographic characteristic of the survey

INSERT TABLE A2 HERE
Table A2. Used ads interest categories

**TABLES**

|  | 1 | 2 | 3 | 4 | 5 | 6 | 7 | Mean | S.E. |
|---|---|---|---|---|---|---|---|---|---|
| Classical music | 30.1% | 19.9% | 15.8% | 12.3% | 10.3% | 3.4% | 8.2% | 3.0 | .16 |
| Electronic music | 13.9% | 18.1% | 13.2% | 13.9% | 16.7% | 11.8% | 12.5% | 3.9 | .16 |
| Hip-hop | 8.2% | 10.3% | 18.5% | 19.2% | 22.6% | 11.0% | 10.3% | 4.1 | .14 |
| Jazz/blues | 12.4% | 24.1% | 15.9% | 20.0% | 11.0% | 9.0% | 7.6% | 3.5 | .15 |
| 'Mulatós' | 29.7% | 11.7% | 13.8% | 14.5% | 9.0% | 11.0% | 10.3% | 3.4 | .17 |
| Pop | 2.1% | 4.8% | 6.9% | 15.2% | 29.0% | 17.2% | 24.8% | 5.2 | .13 |
| Rap | 11.7% | 15.2% | 13.8% | 16.6% | 19.3% | 11.7% | 11.7% | 4.0 | .16 |
| Rock | 13.8% | 13.8% | 13.1% | 11.0% | 15.9% | 9.0% | 23.4% | 4.2 | .18 |
| World-music | 7.8% | 6.4% | 14.2% | 21.3% | 17.7% | 17.0% | 15.6% | 4.5 | .15 |

Table 1. Self-reported music genre preferences in the survey

| Genre | Percent | Valid percent |
|---|---|---|
| Classical music | 2.8% | 3.8% |
| Electronic music | 17.4% | 23.7% |
| Hip-hop/rap | 8.4% | 11.4% |
| Jazz/blues | 2.2% | 3.0% |
| 'Mulatós' | 1.7% | 2.3% |
| Pop | 18.3% | 24.9% |
| Rock | 20.4% | 27.7% |
| World- music | 2.3% | 3.1% |
| *Other music* | 14.6% | |
| *Not music genre* | 12.0% | |

Table 2. Classification of the liked Facebook pages classified as music (N=3803)

|  | Mean | Std error | Max | More than 1 page likes (%) |
|---|---|---|---|---|
| Classical music | .8 | .35 | 51 | 10.7% |
| Electronic music | 6.8 | 2.76 | 398 | 47.3% |
| Hip-hop/rap | 6 | 1.25 | 136 | 42.7% |
| Jazz/blues | .7 | .32 | 47 | 8.0% |
| 'Mulatós' | .5 | .17 | 22 | 10.0% |
| Pop | 13.9 | 2.47 | 230 | 66.7% |
| Rock | 9.8 | 2.39 | 281 | 53.3% |
| World-music | .7 | .29 | 41 | 7.3% |

Table 3. Music genre page likes on respondent level

|  | Classical music | Electronic music | Hip-hop/rap | Jazz/blues | 'Mulatós' | Pop | Rock | World-music |
|---|---|---|---|---|---|---|---|---|
| Classical music | .26** | -.09 | -.21* | .04 | .12 | -.16* | .11 | .06 |
| Electronic music |  | .27** | .20* | -.13 | -.15 | .07 | -.12 | .20* |
| Hip-hop/rap |  |  | .44** | .05 | .32** | .26** | -.09 | .22** |
| Jazz/blues |  |  |  | .15 | -.02 | -.04 | .11 | -.05 |
| 'Mulatós' |  |  |  |  | .23** | .24** | .01 | -.08 |
| Pop |  |  |  |  |  | .21* | -.10 | .14 |
| Rock |  |  |  |  |  |  | .38** | .20* |
| World-music |  |  |  |  |  |  |  | -.02 |

Table 4. Correlation of self-reported interest data (row) and digitally expressed interest data (column)

*p<0.05 **p<0.01*

|  |  | Frequency | Percent |
|---|---|---|---|
| Gender | Male | 37 | 24,7% |
|  | Female | 113 | 75,3% |
| Age category | 18-28 | 95 | 63,3% |
|  | 29-48 | 36 | 24,0% |
|  | 49+ | 19 | 12,7% |
| Education | Less than secondary | 14 | 9,3% |
|  | Secondary | 96 | 64,0% |
|  | University | 40 | 26,7% |

Table A1. Basic socio-demographic characteristic of the survey

| Music genre | Ads interest groups |
|---|---|
| Pop | Pop music |
| Rock | Blues music, Blues-rock, Heavy metal music, Rock music, Rhythm and blues music |
| Electronic | Electro music, Electronic dance music, Electronic music, House music, Psychedelic music, Trance music |
| Hip-hop/rap | Hip-hop music. Rap |
| Classical | Classical music |
| Jazz | Jazz music |
| World music | Country music, Hungarian folk music, Folk music, World music |

Table A2. Used ads interest categories